\documentclass[final]{agujournal}
\usepackage{amssymb}
\usepackage{amsmath}

\journalname{Water Resource Research}

\begin{document}


\title{Rapid Processing of $^{85}$Kr/Kr Ratios using Atom Trap Trace Analysis}

\authors{J. C. Zappala\affil{1,2}, K. Bailey\affil{1}, P. Mueller\affil{1}, T. P. O'Connor\affil{1}, R. Purtschert\affil{3}}

\affiliation{1}{Physics Division, Argonne National Laboratory, Argonne, Illinois 60439, USA}
\affiliation{2}{Department of Physics and Enrico Fermi Institute, University of Chicago, Chicago, Illinois 60637, USA}
\affiliation{3}{Climate and Environmental Physics Division, Physics Institute, University of Bern, CH-3012 Bern, Switzerland}

\correspondingauthor{Jake C. Zappala}{jczappala@uchicago.edu} 


\begin{keypoints}
\item Krypton-85 is a useful tracer for young groundwater in the age range of 5-50 years
\item Throughput of krypton-85 abundance measurements increased twelvefold from previous state-of-the-art
\end{keypoints}


\begin{abstract}
We report a methodology for measuring $^{85}$Kr/Kr isotopic abundances using Atom Trap Trace Analysis (ATTA) that increases sample measurement throughput by over an order of magnitude to 6 samples per 24 hours. The noble gas isotope $^{85}$Kr (half-life = 10.7 yr) is a useful tracer for young groundwater in the age range of 5-50 years. ATTA, an efficient and selective laser-based atom counting method, has recently been applied to $^{85}$Kr/Kr isotopic abundance measurements, requiring 5-10 $\mu$L of krypton gas at STP extracted from 50-100 L of water. Previously a single such measurement required 48 hours. Our new method demonstrates that we can measure $^{85}$Kr/Kr ratios with 3-5\% relative uncertainty every 4 hours, on average, with the same sample requirements.
\end{abstract}


\section{Introduction}

The noble gas isotope $^{85}$Kr is a radioactive nuclide with a half-life of 10.739 $\pm$ 0.014 years \citep{S2014}. It occurs naturally in the atmosphere, produced by cosmic radiation, but at a rate four orders of magnitude lower than current global emission from nuclear fuel reprocessing \citep{A2013}. Due to this anthropogenically increased abundance in the atmosphere and a precise understanding of its input function, $^{85}$Kr can be applied as a tracer to date young groundwater on the order of 5-50 years old.

Tracers in this age regime are crucial to water resource management given the global increased dependency on groundwater, including instances of complete dependency on young, shallow groundwater for drinking water \citep{M2005}. $^{85}$Kr provides an excellent compl\replaced{i}{e}ment for determining ages when taken with other existing tracers in this age regime, such as chlorofluorocarbons (CFCs) \citep{C1995} and $^{3}$H/$^{3}$He, which have both independent input functions \replaced{\citep{L1999} and unrelated corrections \citep{V2007} from $^{85}$Kr}{and corrections from $^{85}$Kr \citep{L1999,V2007}}. Moreover, CFCs are subject to local contamination \citep{P2006} and $^{3}$H/$^{3}$He-dating is highly sensitive to natural degassing \citep{V2007}. In contrast, $^{85}$Kr is steadily released into the atmosphere in a manner that is both monitored and well understood \citep{A2013}, making it a robust tool for dating. \replaced{$^{85}$Kr measurements also have an application towards nuclear non-proliferation efforts given that the isotope is released during nuclear fuel reprocessing efforts \citep{K2010}.}{$^{85}$Kr also has a number of applications beyond groundwater dating, such as monitoring air for nuclear fuel processing activities \citep{K2004, K2010}, monitoring gas transport in the unsaturated zones (which can differ significantly from water transport) \citep{C1995}, and as a tracer of ocean water ventilation and shallow mixing \citep{S1975}.}

As a tracer for dating groundwater, $^{85}$Kr has been successfully applied on many occasions using low-level gas proportional counting (LLC), both on its own \citep{S1992} and in conjunction with other isotopic tracers \added{for deconvolving the age distributions of mixed groundwater} \replaced{\citep{C2007,A2009,M2014,D2014,A2016}}{\citep{C2007,A2009,V2013,M2014,D2014,A2016}}. $^{85}$Kr samples are collected by degassing groundwater samples in the field, and then separating krypton from the bulk gas in the laboratory \citep{P2013}. Recent developments have both decreased seperation times and increased krypton yields of groundwater samples\added{. 10 $\mu$L of krypton gas (STP) can be purified in the laboratory from 10 L of air in approximately 75 minutes. In the field, this amount of air can typically be degassed from 100 L of groundwater in 30-60 minutes} \citep{Y2016}.

However, despite these improvements, $^{85}$Kr-dating has not been applied routinely at a large scale due to the slow processing time and comparatively large sample volume requirements of LLC \citep{P1999,L2005}. The development of Atom Trap Trace Analysis (ATTA), a laser-based atom counting method \citep{C1999}, has sought to provide the necessary tool to make large scale analysis of $^{85}$Kr viable. The ATTA-3 instrument at Argonne National Laboratory (ANL), has been used to routinely measure isotopic abundances of $^{81}$Kr and $^{85}$Kr in groundwater samples \citep{J2012} using 5-10 $\mu$L of krypton gas at STP extracted from 50-100 L of water \added{or 5-10 L of air at STP}. In the past, a single sample measurement required 48 hours.

\indent We report here on a new methodology for $^{85}$Kr analysis through ATTA. We demonstrate that, by using this method on a newly improved ATTA-3 system described in \citet{Z2016}, we now have the ability to continuously measure $^{85}$Kr/Kr ratios with 3-5\% error every 4 hours, on average, increasing the sample throughput by a factor of twelve from \citet{J2012}. We do so with no increase in sample size requirements. We show this method to be linear and repeatable, and present an understanding and control over systematic effects due to cross-sample contamination on the 0.8\% level.

\section{Atom Trap Trace Analysis Method}

The ATTA technique is described fully in \citet{C1999, J2012, Z2016}, but summarized here briefly: krypton gas is injected into a vacuum system and passes through a liquid-nitrogen-cooled, radio-frequency plasma discharge. The cooling slows the atoms and the plasma transfers a fraction of the atoms into a metastable electronic state. From this state, the atoms are resonantly excited with 811 nm lasers used throughout the system to further slow and trap the atoms in a magneto optical trap (MOT). By measuring the loading rates of both the radioactive ($^{81}$Kr and $^{85}$Kr) and stable ($^{83}$Kr) isotopes into the trap, we can obtain an isotopic ratio. Furthermore, in order to remove any systematic effects from changes in efficiency that may occur in the system, we also measure a krypton reference gas immediately after measuring the sample that same day. With the isotopic ratios in both the sample and the reference, we ultimately report a ``superratio" (SR) defined for $^{85}$Kr as
\begin{equation}
\label{SR}
	^{85}\text{Kr}_{\text{SR}} = \frac{^{85}\text{Kr}_{\text{Sample}}/^{83}\text{Kr}_{\text{Sample}}}{^{85}\text{Kr}_{\text{Reference}}/^{83}\text{Kr}_{\text{Reference}}}.
\end{equation}
Such a routine analysis of both the sample and the reference requires 6 hours of total atom trapping. At current, however, the improved ATTA-3 system described in \citet{Z2016} requires 24 hours to complete a single measurement. 16 hours of this time is devoted to ``washing'' the system to decrease cross-sample contamination caused by our plasma discharge, which implants krypton ions from the sample into our vacuum chamber walls. To remove this implanted krypton the plasma discharge is run using argon gas. This process requires that the system return to room temperature to completely remove frozen krypton, meaning 2 hours of the time are spent warming and later re-cooling the liquid-nitrogen source.

However, for measuring only $^{85}$Kr, we can employ a new method. Since $^{85}$Kr/Kr isotopic abundances are 10 times higher than $^{81}$Kr/Kr, a routine measurement would have sufficient statistics to reach a level of 2-3\% error in 0.25 hours, subsequently reducing the amount of krypton being embedded in the system during such a short run. In addition, we can also remove the liquid-nitrogen cooling. This will increase the mean velocity of the atoms, reducing the efficiency of our trap by a factor of 4, lengthening the measurement time to 1 hour; however, it saves 2 hours by removing the heating/cooling cycle.

Here, we present a new measurement procedure without liquid-nitrogen cooling for rapid-processing of $^{85}$Kr/Kr ratios using the ATTA system. First we describe a contamination model for this new method that allows us to control the systematic effects caused from the residual cross-sample contamination. Then we apply that model to six calibration samples measured in a 24-hour period. Samples for this experiment were prepared at the University of Bern and ANL. Their activities were measured using LLC at the University of Bern, and their $^{85}$Kr superratios, defined in equation (\ref{SR}), were measured using the routine ATTA technique that includes liquid-nitrogen cooling and is described at the beginning of this section. These results are reported in first two columns of Table \ref{cal}. \added{The reference gas is represented by the sample J3.}

\begin{table}
\caption{Krypton calibration samples measured using LLC, ATTA, and the new rapid-processing procedure.$^{a}$}
\centering
\begin{tabular}{l c c c c}
\hline
& LLC Activity$^{b}$ &  ATTA $^{85}$Kr$_{\text{SR}}$ & Rapid ATTA & Rapid ATTA \\
& & & $^{85}$Kr$_{\text{SR}}$ (raw) & $^{85}$Kr$_{\text{SR}}$ (corrected)$^{c}$ \\
\hline
J5 & 269 $\pm$ 13 & 8.0 $\pm$ 0.5$^d$ & 7.7 $\pm$ 0.3 & 7.8 $\pm$ 0.3 \\
J4 & 36.2 $\pm$ 3.1 & 1.04 $\pm$ 0.03 & 1.09 $\pm$ 0.05 & 1.09 $\pm$ 0.05 \\
J3 & 32.1 $\pm$ 1.2 & 0.94 $\pm$ 0.01 & 0.95 $\pm$ 0.05 & 0.95 $\pm$ 0.05 \\
J2 & 18.2 $\pm$ 0.6 & 0.53 $\pm$ 0.02 & 0.54 $\pm$ 0.03 & 0.54 $\pm$ 0.03 \\
J1 & 8.9 $\pm$ 0.4 & 0.25 $\pm$ 0.01 & 0.26 $\pm$ 0.02 & 0.25 $\pm$ 0.02 \\
J0 & <1.0 & <0.013 & 0.032 $\pm$ 0.006  & <0.021 \\
& (90\% C.L.) & (90\% C.L.) & &(90\% C.L.) \\
\hline
\multicolumn{5}{l}{$^a$ All ATTA results are expressed using the superratio (SR) defined in equation (\ref{SR}).}\\
\multicolumn{5}{l}{$^b$ Reported in decays per minute per cc of Kr gas at STP, adjusted to 3 March 2016.}\\
\multicolumn{5}{l}{$^c$ Corrected values include adjustment from the contamination model. Raw values do not.}\\
\multicolumn{5}{l}{$^d$ Measured using 1$\mu$L of Kr gas to prevent extensive contamination of $^{85}$Kr.}\\
\end{tabular}
\label{cal}
\end{table}

\begin{figure*}
\centering
\includegraphics[width=33pc]{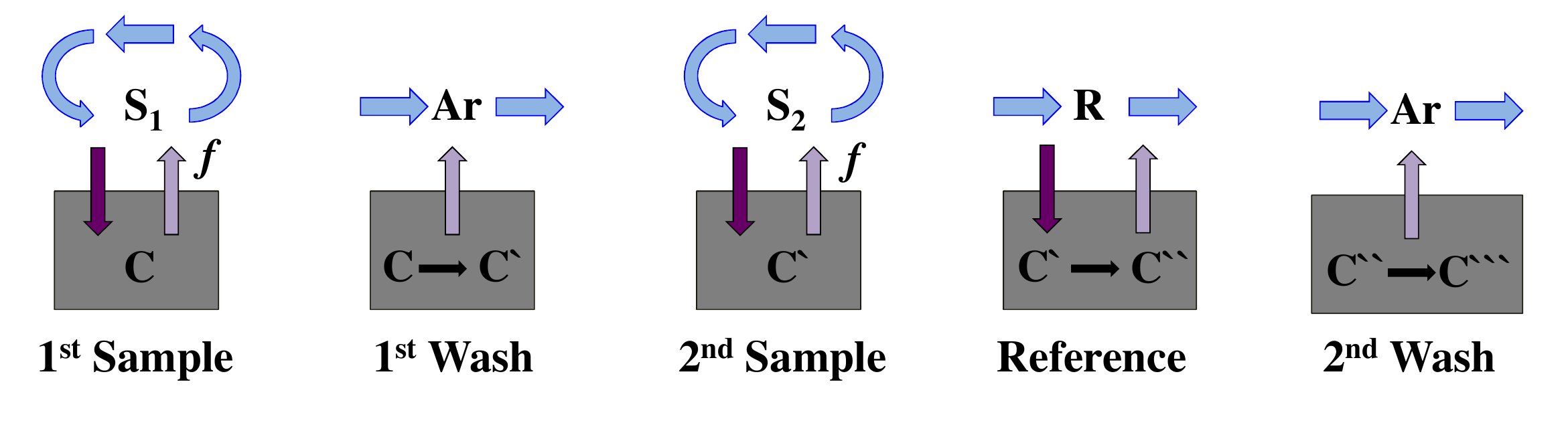}
\caption{\small{A diagram for the sequence S$_1$-S$_2$-R. During sample measurements, there is gas exchange between the contaminant in the chamber wall and the sample gas, which is being recirculated. During washes, argon gas is flowed through the system without recirculation, removing some contaminant. During the reference measurements, reference gas is flowed without recirculation. It enters the chamber walls and reduces the presence of the previous contaminant simultaneously. The contaminant changes in each step as described by our model (see text for details).}}
\label{procedure}
\end{figure*}

For the proceeding sections, we perform all experiments under our liquid-nitrogen free ``rapid-processing'' conditions. Measurements are conducted in the manner illustrated in Figure \ref{procedure}: a sample is measured for $^{85}$Kr/Kr for 1 hour, followed by a 2.25 hour argon wash, another 1 hour sample measurement, a 1 hour reference measurement, and finally another 2.25 hour argon wash before the cycle is repeated. This timing permits us to measure the $^{85}$Kr/Kr ratio of one sample every 4 hours, on average. To define a shorthand for the following sections, a measurement of ``$S_1$-$S_2$-$R$", would mean a measurement of S$_1$ as the first sample and S$_2$ as the second, followed by a reference measurement $R$. During the sample measurements, gas is recirculated in the system (as it is for a typical ATTA measurement) due to the small size of the samples. During the washes and the reference measurements, the gas is flowed continuously and discarded. \added{Due to this systematic difference between measurements made with and without gas recirculation, the $^{85}$Kr superratio of our reference gas measured as a sample (J3) on ATTA is 0.94. However, this systematic difference is found to be consistent and is taken into account as a constant calibration factor. Thus, it does not effect the linearity or reliability of our measurements.}

\section{Contamination Model}

\added{Due to the significantly reduced wash times in this rapid-processing procedure, it is crucial that we develop a model for the cross-sample contamination effects on our system. The goal of this section is to develop a relatively simple but reliable empirical model that quantitatively describes the data without going into the complexities of the cross-contamination mechanism such as specific implantation sites and chamber volumes.}

Following the diagram in Figure \ref{procedure}, we\deleted{ can} consider the collective surfaces of ATTA-3 affected by implantation and the volume of sample gas to be two distinct reservoirs. The former is filled with contaminant from previous samples and the latter is filled with our sample to be measured. The contaminant has its own $^{85}$Kr/Kr ratio, which we define as $C$. Due to the plasma discharge there is an exchange: sample gas enters the surface reservoir and contaminant leaks into the volume of the sample gas. The contamination that leaks into the sample becomes part of our measured value. The sample which enters into the surfaces replaces some\deleted{ fixed} fraction $x$ of the current contaminant in the reservoir, reducing the influence of each previous sample's contribution to the contaminant by some fraction 1-$x$. \replaced{The wash afterward then simply reduces all contributions in proportion}{The wash procedure afterward reduces the overall number of contaminating krypton atoms by replacing them with argon atoms. However, the wash does not alter the $^{85}$Kr/Kr ratio of the contaminant since it affects all implanted krypton isotopes equally}.

Thus, if there is some contaminant $C$ before $S_1$ is measured, then after the wash we now have a contaminant
\begin{equation}
\label{def}
C' = C(1-x) + S_1x
\end{equation}
We test such a model by attempting to find a repeatable value for $x$. \added{This $x$ is particular to our current vacuum system and will require reevaluation if changes are made to the chamber.}

To find $x$ we first need to know how much contamination we have in our system. We define $f$, the average portion of the sample ($S$) volume that is replaced by the contaminant ($C$) during a measurement of the sample ($M_S$) as 
\begin{equation}
\label{frac}
M_{S} = (1-f)S + fC \quad\quad\text{where}\quad\quad f=\frac{\frac{1}{2}tR_{\text{Kr}}}{P_{\text{avg}}}.
\end{equation}
Here $t$ is the length of the measurement, $R_{\text{Kr}}$ is the linear outgassing rate of the contaminant, and $P_{\text{avg}}$ is the\added{ average partial} pressure of\added{ krypton during} the run. \added{The factor $\frac{1}{2}tR_{\text{Kr}}$ gives us the integrated contamination injected into the gas volume, which is normalized by the partial pressure of krypton, $P_{\text{avg}}$.} The only unknown here is the outgassing rate. To determine the rate we regularly conduct an outgassing test prior to each measurement: the system is filled with argon gas and the gas is recirculated with the plasma active. We then measure how much krypton leaches out of the wall over a few minutes using a SRS Residual Gas Analyzer and extrapolate a krypton outgassing rate due to the argon discharge, $R_{\text{Ar}}$. However, we need to determine the krypton outgassing rate in the presence of a krypton discharge, which should be proportional, but not equal to the rate we have measured, i.e. $R_{\text{Kr}} = b R_{\text{Ar}}$.

To determine $b$, we first clean the system for longer than the normal wash period such that the outgassing rate is more than a factor of 4 lower than the typical rates we expect in these measurements. In this ``clean start,'' if we measure $S_1$-$S_2$-$R$, then the contaminant $C \approx S_1$ when we measure $S_2$. Accordingly we obtain
\begin{equation}
M_{S_2} = (1-f)S_2 + fC = (1-f)S_2 + fS_1\nonumber
\end{equation}
Using our calibration samples, we start with a clean system and then measure J5-J0-R. This simplifies the above equation even further, since J0 is devoid of $^{85}$Kr ("$^{85}$Kr-dead") and thus $S_2=0$. With only the second term, we can solve for $b$. We used three such measurements to determine that $b$ = 2.4 $\pm$ 0.3.

Now that we know the value of $f$, we can work to find $x$ by applying the model. If we consider measuring $S_1$-$S_2$-$R$-$S_3$ with our clean start, the model gives us the following for the third sample measurement 
\begin{eqnarray}
\label{contam2}
M_{S_3} &=& S_3(1-f) + fC''\nonumber\\ &=& S_3(1-f) + f(C'(1-x) + Rx)\nonumber\\ &=& S_3(1-f) + f(S_1(1-x) + S_2x)(1-x) + Rx)\nonumber
\end{eqnarray}
From here, we can solve for $x$. We ran two separate sequences, J2-J5-R-J0 and J5-J0-R-J0-J0, to solve for $x$ and found that $x$ = 0.60 $\pm$ 0.02. Note that we have considered the reference to both be a sampling \emph{and} a wash procedure. Yet, despite it only being 1 hour instead of 2.25, we still found consistent results. The reason is that, as shown by solving for $b$, a krypton wash is $\sim$2.4 times more effective at extracting krypton than an argon wash. \added{As such, we could increase the efficiency of the wash by using $^{85}$Kr-dead krypton gas as our wash gas. However, sufficient amounts of $^{85}$Kr-dead krypton gas are not readily available.}

With this repeatable value for $x$ we have determined a simple and consistent model for describing our contamination in this rapid-processing mode.

\section{Rapid Processing Results}

\begin{figure}
\centering
\includegraphics[width=23pc]{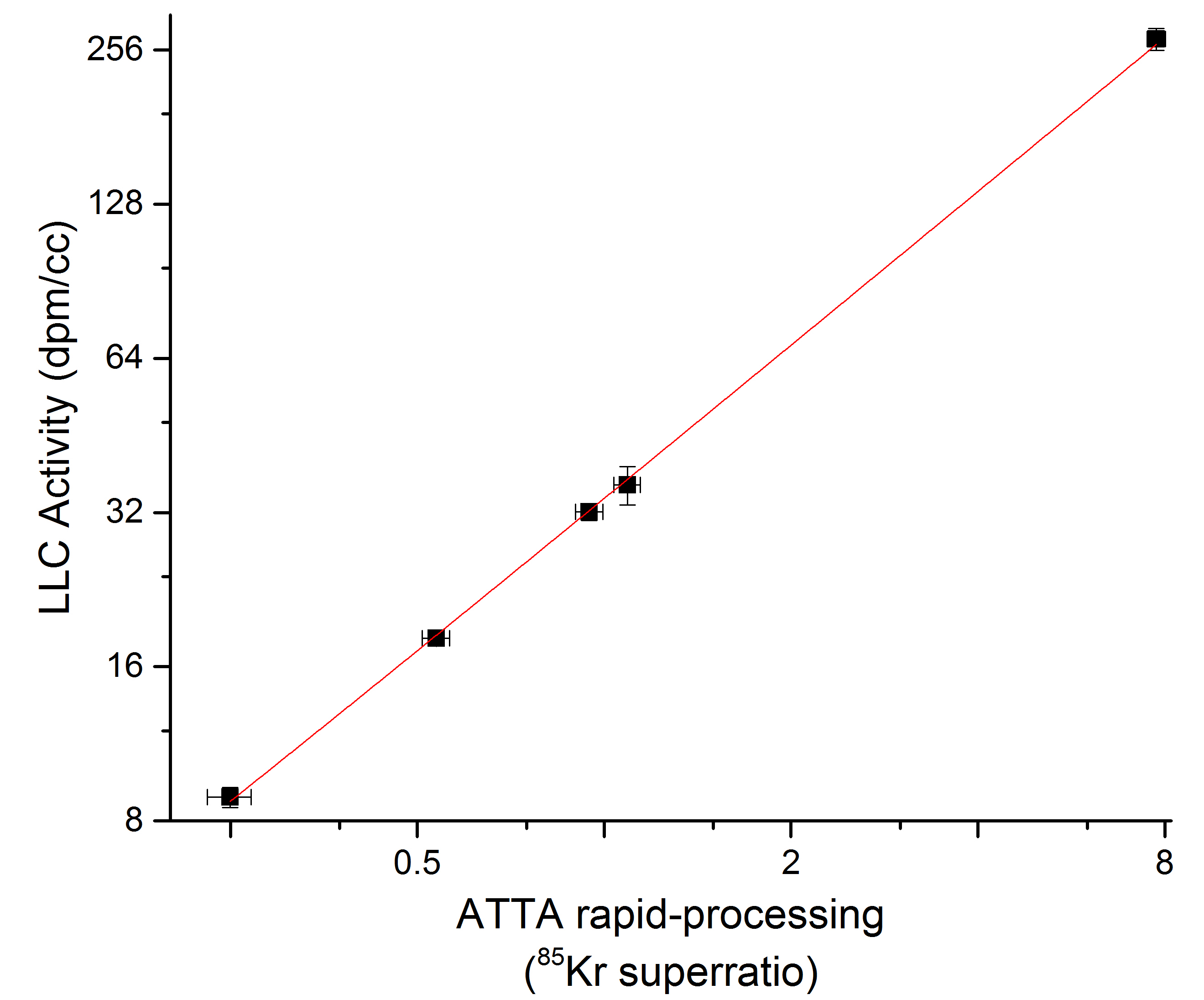}
\caption{\small{A demonstration of linearity for rapid-processing superratio measurements on ATTA via comparison with LLC results. Both axes are drawn on a log$_2$ scale. J0 was measured to have a superratio of <0.021 (90\% C.L.), but is not shown due to the log scale. The fit has a reduced chi-squared of 0.2.}}
\label{line}
\end{figure}

\begin{figure}
\centering
\includegraphics[width=30pc]{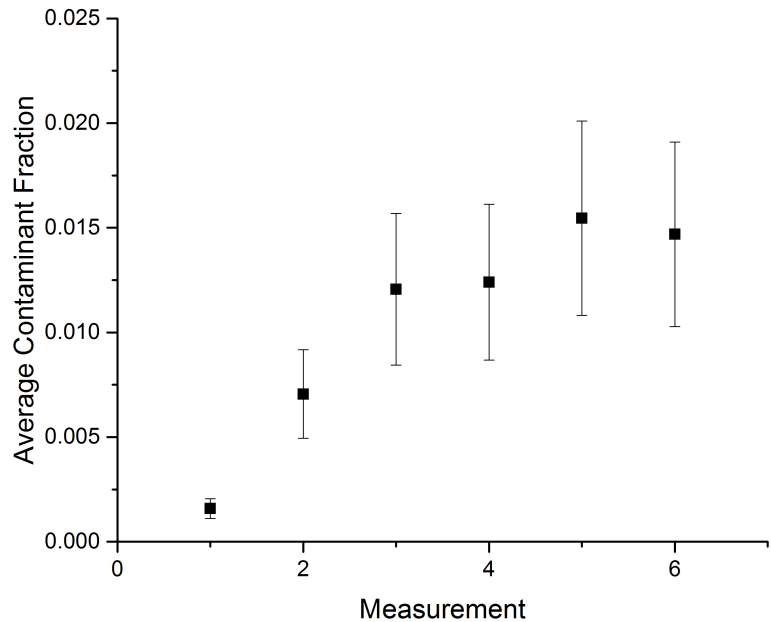}
\caption{\small{Average contamination during each measurement over six sequential measurements during a 24-hour period.}}
\label{con}
\end{figure}

We measured six calibration samples in a 24-hour period (measured in the order J2-J5-R-J0-J4-R-J1-J3-R). The $^{85}$Kr superratios determined from these measurements are listed in Table \ref{cal} in the third column, and listed with corrections from the contamination model in the fourth column. The LLC activities of the samples are plotted against these corrected values in Figure \ref{line}\deleted{Previous caption: A demonstration of linearity for rapid-processing superratio measurements on ATTA via comparison with LLC results. Both axes are drawn on a log$_2$ scale. J0 was measured to have a superratio of <0.021 (90\% C.L.), but is not shown or applied to the fit. Instead the fit is forced through zero. The fit has a reduced chi-squared of 0.27.} and fit to a line. The measurement of J0 does not appear in the figure \replaced{since it is reported as a limit, however, the fit shown in the figure is forced through zero in recognition that both our rapid-processing method and LLC yield below detection limit results after correction}{due to the log$_2$ scaling, but is included in the fit}. The reduced chi-squared of the fit is \replaced{0.27}{0.2} We also see that in the six samples we measured the contamination fraction per sample saturated below the 2.5\% level, as seen in Figure \ref{con}. Based on this saturation level and the errors of our contamination model, the correction will add a maximum error of below 0.8\% to the $^{85}$Kr/Kr isotopic abundance measurements, which typically have 3-5\% statistical error.

The rapid-processing measurements' agreement with typical ATTA measurements and their linear relationship with LLC activities, demonstrates the validity of this approach. This method increases the throughput of $^{85}$Kr/Kr isotopic abundance measurements on a single ATTA system by a factor of twelve. The agreement of this calibration over such a large range of activities (J5 being \replaced{up to 3-5 times the typical value in the atmosphere}{nearly 4 times higher than the typical 75 dpm/cc activity in the atmosphere of the northern hemisphere \citep{A2013}}) also shows that our contamination model can even handle enrichment levels we would normally wish to avoid in the standard ATTA system. With this rapid-processing procedure validated, ATTA is ready to increase the capacity for $^{85}$Kr-dating in the geoscience community.


\acknowledgments
We thank Zheng-Tian Lu for his comments and suggestions. This work is supported by Department of Energy, Office of Nuclear Physics, under Contract No. DEAC02-06CH11357. We also acknowledge funding from an Argonne/University of Chicago Collaborative Seed Grant. Interested readers can access the data reported in this paper by directly contacting the corresponding author.


\listofchanges

\end{document}